%% file: paper.tex
\documentclass{jfm}

\usepackage{graphicx}
\usepackage{natbib}
\usepackage[normalem]{ulem}

\usepackage{color} % MK added
\usepackage{comment} % MK added
\usepackage{amsmath} % MK added
\usepackage[textsize=scriptsize]{todonotes} % MK added
\newcommand{\karman}{von K\'{a}rm\'{a}n } % MK added

\ifCUPmtlplainloaded \else
  \checkfont{eurm10}
  \iffontfound
    \IfFileExists{upmath.sty}
      {\typeout{^^JFound AMS Euler Roman fonts on the system,
                   using the 'upmath' package.^^J}%
       \usepackage{upmath}}
      {\typeout{^^JFound AMS Euler Roman fonts on the system, but you
                   dont seem to have the}%
       \typeout{'upmath' package installed. JFM.cls can take advantage
                 of these fonts,^^Jif you use 'upmath' package.^^J}%
      }
  \else
  \fi
\fi

\ifCUPmtlplainloaded \else
  \checkfont{msam10}
  \iffontfound
    \IfFileExists{amssymb.sty}
      {\typeout{^^JFound AMS Symbol fonts on the system, using the
                'amssymb' package.^^J}%
       \usepackage{amssymb}%
       \let\le=\leqslant  \let\leq=\leqslant
         
      }{}
  \fi
\fi

\ifCUPmtlplainloaded \else
  \IfFileExists{amsbsy.sty}
    {\typeout{^^JFound the 'amsbsy' package on the system, using it.^^J}%
     \usepackage{amsbsy}}
    {\providecommand\boldsymbol[1]{\mbox{\boldmath $##1$}}}
\fi

\newsavebox{\astrutbox}
\sbox{\astrutbox}{\rule[-5pt]{0pt}{20pt}}

\title[DNS of turbulent channel flow up to $Re_\tau \approx$ 5200]{Direct numerical simulation of turbulent channel flow up to $Re_\tau \approx $ 5200}

\author[M. Lee and R. D. Moser]%
{Myoungkyu Lee$^1$ and Robert D. Moser$^{1,2}$\thanks{Email address for correspondence: rmoser@ices.utexas.edu}}

\affiliation{$^1$Department of Mehcnanical Engineering, The University of Texas at Austin, TX 78712, USA\\[\affilskip]
$^2$Center for Predictive Engineering and Computational Sciences, Insititute for Computational Engineering and Sciences, The University of Texas at Austin, TX 78712, USA}

\pubyear{2010}
\volume{650}
\pagerange{119--126}
\date{?; revised ?; accepted ?. - To be entered by editorial office}
\begin{document}

\maketitle

\begin{abstract}
A direct numerical simulation of incompressible channel flow at
$Re_\tau=5186$ has been performed, and the flow exhibits a number of the
characteristics of high Reynolds number wall-bounded turbulent flows.
For example, a region where the mean velocity has a logarithmic
variation is observed, with \karman constant $\kappa =
0.384 \pm 0.004$. There is also a logarithmic dependence of the variance
of the spanwise velocity component, though not the streamwise component.
A distinct separation of scales exists between the large outer-layer
structures and small inner-layer structures. At intermediate
distances from the wall, the one-dimensional spectrum of the streamwise
velocity fluctuation in both the streamwise and spanwise directions
exhibits $k^{-1}$ dependence over a short range in $k$. 
Further, consistent with previous experimental
observations, when these spectra are multiplied by $k$ (premultiplied
spectra), they have a bi-modal structure with local peaks located at
wavenumbers on either side of the $k^{-1}$ range.
\end{abstract}

\begin{keywords}
%Authors should not enter keywords on the manuscript, as these must be chosen by the author during the online submission process and will then be added during the typesetting process (see http://journals.cambridge.org/data/\linebreak[3]relatedlink/jfm-\linebreak[3]keywords.pdf for the full list)
\end{keywords}

\section{Introduction}
\label{sec:intro}
\input{intro}

%\newpage

\section{Simulation Details}
\label{sec:method}
\input{method}

%\newpage

\section{Results}

In the discussion to follow, when comparing
results from different simulations, data from the three cases with the
highest Reynolds number (LM5200, LJ4200 and BPO4100) are plotted with
solid lines while the lower Reynolds number cases are with dashed
lines. The experimental data are plotted with symbols. The legend of
line styles, colors and symbols is given in
Table~\ref{table:simulation_parameters}.

\label{sec:results}
\subsection{Mean velocity profile}
\label{sec:mean}
\input{mean_velocity_profile}

%\newpage

\subsection{Reynolds Stress Tensor}
\label{sec:fluctuations}
\input{velocity_fluctuation}
%\newpage

\subsection{Discrepancies Between Simulations}
\label{sec:resolution}
\input{resolution}

\subsection{Energy spectral density}
\input{energy_spectra}

%\newpage

%\subsection{Total stress}
%\input{total_stress}
%\newpage

\section{Discussion and Conclusion}
\label{sec:conclusion}
\input{conclusions}

%\newpage

\section{Acknowledgment}

The work presented here was supported by the National Science
Foundation under Award Number [OCI-0749223] and the Argonne Leadership
Computing Facility at Argonne National Laboratory under Early Science
Program(ESP) and Innovative and Novel Computational Impact on Theory
and Experiment Program(INCITE) 2013. We wish to thank N. Malaya and R. 
Ulerich for software engineering assistance. We
are also thankful to M. P. Schultz and K. A. Flack for providing their
experimental data.

%\newpage
%\appendix
%\section{Uncertainty estimation}
%\input{appendix}

% susie put cite commands here, don't bother with citet etc just yet.

\bibliographystyle{jfm}
% Note the spaces between the initials

%\bibliography{/workspace/mendeley_library/bibtex/library}
\bibliography{reference}

%\bibliography{jfm-instructions}

\end{document}

%% file: intro.tex
Recently, relatively ``high'' Reynolds number wall-bounded turbulence
has been investigated with the help of innovations in measurement
technologies \citep{Nagib2004,Kunkel2006,Westerweel2013,Bailey2014}
and computation \citep{Lee2013,Borrell2013,ElKhoury2013}. Because of
the simplicity in geometry and boundary conditions, pressure-driven
turbulent flow between parallel walls (channel flow) is an excellent
vehicle for the study of wall-bounded turbulence via direct numerical
simulation (DNS).  Since \citet{Kim1987a} showed agreement between DNS
and experiments in channel flow in 1987, DNS of channel flow has been
used to study wall-bounded turbulence at ever higher Reynolds number,
as advances in computing power allowed. For example, channel flow DNS
were performed at $Re_\tau = 590$ \citep{Moser1999a}, $Re_\tau =
950$ \citep{DelLamo2004}, and $Re_\tau = 2000$ \citep{Hoyas2006}. More
recently, simulations with $Re_\tau \approx 4000$ were performed
separately by
\citet{Lozano-Duran2014} in a relatively small domain size and by \citet{Bernardini2014}.
Also, to investigate the differences between channel flow turbulence
and boundary layer turbulence, a simulation at $Re_\tau = 2000$ of a
zero pressure gradient boundary layer was performed
by \citet{Sillero2013}.

The idealized channel flow that is so straight-forward to simulate is
more difficult to realize experimentally because of the need for side
walls. Measurements of channel flow are not as numerous as for other
wall bounded turbulent flows, but a number of studies have been
conducted over the years
 \citep{Comte-Bellot1963,Dean1976,Johansson1982,Wei1989,Zanoun2003,Zanoun2009,Monty2009}.
%\todo{RDM: There were also simulations by Christensen, MK: I checked
%  the paper Natrajan and Christensen POF 2006. It is about ZPGBL. I
%  don't think this is the right ref in this particular context.}
Of particular interest here will be the channel measurements made by
\citet{Schultz2013} at Reynolds number up to $Re_\tau = 6000$, because
they bracket the Reynolds number simulated here.

Turbulent flows with $Re_\tau$ on the order of $10^3$ and greater are
of interest because this is the range of Reynolds numbers relevant to
industrial applications \citep{Smits2013}. Further, it is in this range
that characteristics of wall-bounded turbulence associated with high
Reynolds number are first manifested \citep{Marusic2010b}, and so
studies of these phenomena must necessarily be conducted at these
Reynolds numbers.  The best-known feature of high-Reynolds number
wall-bounded turbulence is the logarithmic law in the mean velocity,
which has been known since the 1930's \citep{Millikan1938}.
The \karman constant $\kappa$ that appears in the log-law, is also an
important parameter for calibrating turbulence
models \citep{Durbin2010,Marusic2010b,Smits2011,SPALART1992}. It was
considered universal in the past, but  \citet{Nagib2008} showed that
$\kappa$ can be different in different flow
geometries. Further, \citet{Jimenez2007} found that
channel flows with $Re_\tau$ up to 2000, do not exhibit a region where
the mean velocity profile strictly follows a logarithmic law, and
showed that a finite Reynolds number correction like that introduced
by \citet{Afzal1976} was more consistent with data in this Reynolds
number range. Similarly, \citet{Mizuno2011} used a different finite
Reynolds number correction to represent the overlap region in boundary
layers, channels and pipes.

For turbulent intensities, \citet{Townsend1976} predicted that, in
high Reynolds number flows, there are regions where the variance of
the streamwise and spanwise velocity components decrease
logarithmically with distance from the wall. This has been observed
experimentally for streamwise
\citep{Kunkel2006,Winkel2012,Hultmark2012,Hutchins2012,Marusic2013}
and spanwise components \citep{Fernholz1996,Morrison2004}, but has
only been observed in the spanwise component in DNS
\citep{Hoyas2008,Sillero2013}. Further, the peak of the streamwise
velocity variance has a weak Reynolds number dependence when it is
scaled by friction velocity, $u_\tau$ \citep{DeGraaff2000,Hoyas2006}.

There has also been recent interest in the role of large scale motions
(LSM) in high Reynolds number wall-bounded turbulent
flow \citep{Kim1999,Hutchins2007,Wu2012}. According to the scaling
analysis of \citet{Perry1986} there is a region where the
one-dimensional spectral energy density has a $k_x^{-1}$
dependence. This has been supported by experimental
evidence \citep{Nickels2005,Nickels2007}, but not numerical simulations.

In summary, there are many characteristics of high-Reynolds-number
wall-bounded turbulence that have been suggested by theoretical
arguments and corroborated experimentally, but which have not been
observed in direct numerical simulations, presumably due to the
limited Reynolds numbers of the simulations. This is unfortunate
because such simulations could provide detailed information about
these high Reynolds number phenomena that would not otherwise be
available.  The direct numerical simulation reported here was
undertaken to address this issue, by simulating a channel flow at
sufficiently high Reynolds number and with sufficiently large spatial
domain to exhibit characteristics of high-Reynolds number turbulence
like those discussed above. The simulation was performed for
$Re_\tau \approx 5200$ with the same domain size used by \citet{Hoyas2006} 
in their $Re_\tau=2000$ simulation.

This paper is organized as follows.  First, the simulation methods and
parameters are described in \S\ref{sec:method}, along with
other simulations and experiments used for comparison. Results of the simulation
that arise due to the relatively high Reynolds number of the
simulation are presented in
\S\ref{sec:results}. Finally, conclusions are offered in
\S\ref{sec:conclusion}.

%% file: method.tex
In the discussion to follow, the streamwise, wall-normal and spanwise
velocities will be denoted $u$, $v$ and $w$ respectively with the mean
velocity indicated by a capital letter, and fluctuations by a
prime. Furthermore, $\langle\cdot\rangle$ indicates the expected value
or average. Thus $U=\langle u\rangle$ and $u=U+u'$. 

The simulations reported here are DNS of incompressible turbulent flow
between two parallel planes. Periodic boundary conditions are applied
in the streamwise ($x$) and spanwise ($z$) directions, and
no-slip/no-penetration boundary conditions are applied at the
wall. The computational domain sizes are $L_x=8\pi\delta$ and
$L_z=3\pi\delta$, where $\delta$ is the channel half-width, so the
domain size in the wall-normal ($y$) direction is $2\delta$. The flow
is driven by a uniform pressure gradient, which varies in time to
ensure that the mass flux through the channel remains constant.

A Fourier-Galerkin method is used in the streamwise and spanwise
directions, while the wall-normal direction is represented using
a B-spline collocation method \citep{Kwok2001,Botella2003}. The
Navier-Stokes equations are solved using the method of
\citet{Kim1987a}, in which equations for the wall-normal
vorticity and the Laplacian of the wall-normal velocity are
time-advanced.  This formulation has the advantage of satisfying the
continuity constraint exactly while eliminating the pressure.  A
low-storage implicit-explicit scheme \citep{Spalart1991} based on
third-order Runge-Kutta for the non-linear terms and Crank-Nicolson
for the viscous terms is used to advance in time. The time step is
adjusted to maintain an approximately constant CFL number of one. A
new highly optimized code was developed to solve the Navier-Stokes
equations using these methods on advanced peta-scale computer
architectures. For more details about the code, the numerical methods,
and how the simulations were run, see \citet{Lee2013,Lee2014}.

The simulations performed here were conducted with resolution
comparable to that used in previous high Reynolds number channel flow
simulations, when measured in wall units. Normalization in wall units,
that is, with kinematic viscosity $\nu$ and friction velocity
$u_\tau$, is indicated with a superscript ``+''. The friction velocity is 
$u_\tau=\sqrt{\tau_w/\rho}$, where $\tau_w$ is the mean wall shear
stress.  For the highest Reynolds number simulation reported here,
which is designated LM5200, $N_x=10240$ and $N_z=7680$ Fourier modes
were used to represent the streamwise and spanwise direction, which
results in an effective resolution $\Delta x^+=L_x^+/N_x=12.7$ and
$\Delta z^+=L_z^+/N_z = 6.4$. In the wall normal direction, the
seventh order B-splines are defined on a set of $1530=N_y - 6$ knot
points ($N_y$ is the number of B-spline basis functions), which are
distributed uniformly in a mapped coordinate $\xi$ that is related to
the wall-normal coordinate $y$ through
\[
\frac{y}{\delta} = \frac{\sin\left(\eta \xi \pi/2 \right)}{\sin\left(\eta \pi/2 \right)},\quad -1\leq \xi \leq 1
\]
The single parameter $\eta$ in this mapping controls how strongly the
knot points are clustered near the wall, with the strongest clustering
occurring when $\eta=1$. In LM5200, $\eta=0.97$, resulting in the
first knot point from the wall at $\Delta y^+_w=0.498$ and the
centerline knot spacing of $\Delta y_c^+=10.3$. The $N_y$ collocation
points are determined as the Greville abscissae \citep{Johnson2005}, in
which, for $n$-degree splines, each collocation point is the average
of $n$ consecutive knot points ($n=7$ here), with the knots at the
boundary given a multiplicity of $n-1$. As a result, the
collocation points are more clustered near the wall than the knot
points. Resolution parameters for all the simulations discussed here
are provided in Table~\ref{table:simulation_parameters}.

Because the mass flux in the channel remains constant, the bulk
Reynolds number $Re_b$ can be specified directly for a simulation,
where $Re_b=U_b\delta/\nu$,
$U_b=\frac{1}{2\delta}\int_{-\delta}^\delta U(y)\,dy$ and $U(y)$ is
the mean streamwise velocity. Four simulations at four different
Reynolds numbers were conducted. Of most interest here is the highest
Reynolds number case, LM5200, for which the bulk Reynolds number is
$Re_b=1.25\times 10^5$, and the friction Reynolds number is
$Re_\tau=5186$ ($Re_\tau=u_\tau\delta/\nu$). The three other cases
simulated, LM180, LM550 and LM1000, were performed for convenience to
regenerate data for previously simulated
cases \citep{Kim1987a,Moser1999a,DelLamo2004} using the numerical
methods used here. The simulation details for each case are summarized
in Table~\ref{table:simulation_parameters}.

In addition, channel flow data from four other sources in the
literature are included here for comparison. First is a simulation at
$Re_\tau=2000$ conducted by \citet{Hoyas2006} 
(HJ2000), which used the same domain size and similar numerical scheme
as the current simulations, differing only in the use of high-order
compact finite differences in the wall-normal direction, rather than
B-splines. A second simulation (LJ4200) by \citet{Lozano-Duran2014} is at $Re_\tau=4179$ and uses the
same numerical methods as HJ2000, but the domain size in
$x$ and $z$ is much smaller. The third simulation (BP04100), which was
done by \citet{Bernardini2014}, is at $Re_\tau=4079$ and
used a domain size not much smaller than that used here, but these
simulations were performed using second-order finite
differences. Finally, experimental data from laser Doppler velocimetry
measurements at two Reynolds numbers ($Re_\tau=4048$ and 5895, SF4000
and SF6000, respectively) are reported by \citet{Schultz2013} and are also included here for
comparison. Summaries of all these data sources are also included in
Table~\ref{table:simulation_parameters}.

\begin{table}
  \begin{center}
  \def~{\hphantom{0}}
  \resizebox{\textwidth}{!}{
  \begin{tabular}{lrrccccccccccl} 
  Name    & $Re_\tau$ & $Re_b$ & Method & $L_x/\delta$ & $L_z/\delta$ & $\Delta x^+$ & $\Delta z^+$ & $\Delta  y^+_{w}$ & $\Delta  y^+_{c}$ & $N_y$ &  $T u_\tau / \delta$ & \multicolumn{2}{c}{Line \& Symbols} \\ [3pt]
  LM180   & 182  & 2,857   & SB  & 8$\pi$ & 3$\pi$ & ~4.5 & 3.1 & 0.074 & ~3.4 & ~192 & 31.9~ & {\color{magenta}$---$}    & (Magenta) \\
  LM550   & 544  & 10,000  & SB  & 8$\pi$ & 3$\pi$ & ~8.9 & 5.0 & 0.019 & ~4.5 & ~384 & 13.6~ & {\color{blue}$---$}       & (Blue)    \\
  LM1000  & 1000 & 20,000  & SB  & 8$\pi$ & 3$\pi$ & 10.9 & 4.6 & 0.019 & ~6.2 & ~512 & 12.5~ & {\color{red}$---$}        & (Red)     \\ 
  LM5200  & 5186 & 125,000 & SB  & 8$\pi$ & 3$\pi$ & 12.7 & 6.4 & 0.498 & 10.3 & 1536 & ~7.80 & {\bf------}               & (Black)   \\
  HJ2000  & 2003 & 43,650  & SC  & 8$\pi$ & 3$\pi$ & 12.3 & 6.1 & 0.323 & ~8.9 & ~633 & 11.~~ & {\color{green}$---$}      & (Green)   \\ 
  LJ4200  & 4179 & 98,302  & SC  & 2$\pi$ & $\pi$  & 12.8 & 6.4 & 0.314 & 10.7 & 1081 & 15.~~ & {\color{blue}{\bf------}} & (Blue)    \\ 
  BPO4100 & 4079 & 95,667  & FD  & 6$\pi$ & 2$\pi$ & ~9.4 & 6.2 & 0.010 & 12.5 & 1024 & ~8.54 & {\color{red}{\bf------}}  & (Red)     \\ 
  SF4000  & 4048 & 94,450  & LDV & ...    & ...    & ...  & ... & ...   & ...  & ...  &   ... & \multicolumn{2}{c}{$\triangle$}       \\ 
  SF6000  & 5895 & 143,200 & LDV & ...    & ...    & ...  & ... & ...   & ...  & ...  &   ... & \multicolumn{2}{c}{$\Diamond$}        \\ 
  \end{tabular}
  }
  \caption{Summary of simulation parameters. $\Delta x$ and $\Delta z$ are in terms of Fourier modes for spectral methods. 
  $\Delta y_w$ and $\Delta y_c$ are grid spacing at wall and center
  line, respectively. $\delta$ - Channel half width, $Re_\tau =
  u_\tau \delta/\nu$, $Tu_\tau/\delta$ - Total simulation time without
  transition, SB - Spectral/B-Spline, SC - Spectral/Compact finite
  difference, FD - Finite difference, LDV - Laser Doppler velocimetry}
  \label{table:simulation_parameters}
  \end{center}
\end{table}

\begin{figure}
\centering
	\includegraphics[width=\textwidth]{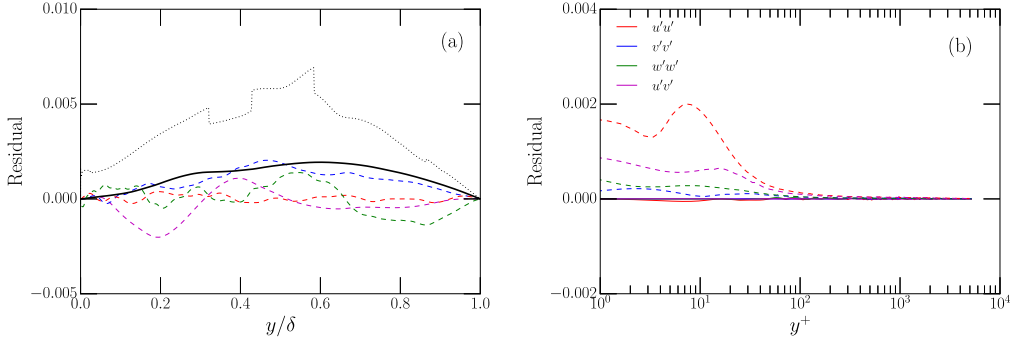}
\caption{Statistical stationary of simulation: (a) Residual in
(\ref{eq:total_stress}), with linestyle legend given in
Table~\ref{table:simulation_parameters} and the dotted line is the standard deviation of
estimated total stress in LM5200; (b) Residual in (\ref{RSTE}) of
LM5200 (solid), and the standard deviation of the estimated statistical error in the sum
of the RHS terms in (\ref{RSTE}) (dashed).}
\label{fig:stationary}

\end{figure}

We used the method described by \citet{Oliver2014} to estimate
the uncertainty in the statistics reported here due to sampling
noise. For LM5200, the estimated standard deviation of the mean velocity
is less than 0.2\% and the estimated standard deviation of
the variance and covariance of the velocity components is less than 0.5\% in
the near-wall region ($y^+ < 100$ say) and 3\% in the outer-region ($y >
0.2\delta$ say).  We also used the total stress and the Reynolds stress
transport equations to test whether the simulated turbulence is
statistically stationary. In a statistically stationary turbulent channel, the total
stress, which is the sum of Reynolds stress and mean viscous stress, is
linear due to momentum conservation:
\begin{equation}
\frac{\partial U^+}{\partial y^+} - \left< u^{\prime }v^\prime \right>^+  \approx 1 - \frac{y}{\delta}
\label{eq:total_stress}
\end{equation}
As shown in fig~\ref{fig:stationary}a, the discrepancy between the
analytic linear profile and total stress profile from the simulations
is less than 0.002 (in plus units) in all simulations, and it is much
smaller than the standard deviation of the estimated total stress of
LM5200. 

The Reynolds stress transport equations, which govern the evolution of
the Reynolds stress tensor, are given by:
\begin{eqnarray}
  \frac{D \left< u_i^{\prime} u_j^{\prime} \right>}{D t} & = &  -\left(\left<u_i^{\prime}u_k^{\prime}\right>\frac{\partial U_j}{\partial x_k} 
  + \left<u_j^{\prime}u_k^{\prime}\right>\frac{\partial U_i}{\partial x_k} \right) - \frac{\partial \left< u_i^{\prime} u_j^{\prime} u_k^{\prime}\right>}{\partial x_k} 
  + \nu \frac{\partial^2 \left< u_i^{\prime} u_j^{\prime}\right>}{\partial x_k \partial x_k} \nonumber \\ 
  && +  \left< p^{\prime}\left(\frac{\partial u_i^{\prime}}{\partial x_j} + \frac{\partial u_j^{\prime}}{\partial x_i}\right)\right> 
  -\left(\frac{\partial \left<p^{\prime}u_i^{\prime}\right>}{\partial x_j}+\frac{\partial \left<p^{\prime}u_j^{\prime}\right>}{\partial x_i}\right) 
  -  2\nu\left< \frac{\partial u_i^{\prime}}{\partial x_k}\frac{\partial u_j^{\prime}}{\partial x_k}\right>
  \label{RSTE}
\end{eqnarray}
While not reported here, all the terms on the right hand side of
(\ref{RSTE}) have been computed from our simulations and the data are
available at {\tt http://turbulence.ices.utexas.edu}.  In a
statistically stationary channel flow, the substantial derivative on
the left of (\ref{RSTE}) is zero.  Hence, any deviation from zero of
the sum of the terms on right hand side of (\ref{RSTE}) is an
indicator that the flow is not stationary. The residual of
(\ref{RSTE}) is shown in fig~\ref{fig:stationary}b (solid lines) in
wall units. The values for all components of the Reynolds stress are
much less than 0.0001 in wall units, which is less than 0.01\% error in
the balance near the wall.  The relative error in the balance
increases to order 1\% away from the wall, as the magnitude of the
terms in (\ref{RSTE}) decrease. Across the entire channel, the
estimated standard deviation of the statistical noise (dashed lines) is much
larger than these discrepancies.

%% file: mean_velocity_profile.tex
\begin{figure}
\centerline{\includegraphics[width=.55\textwidth]{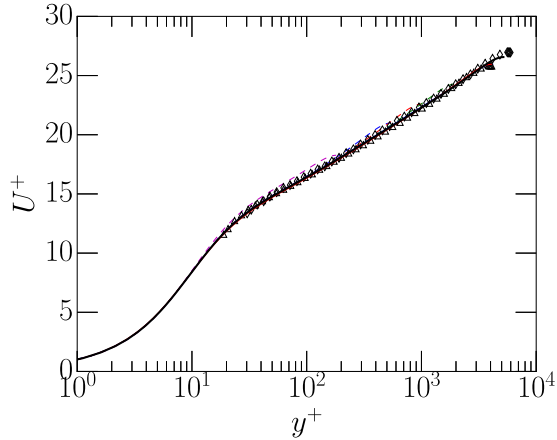} }
\caption{Mean streamwise velocity profile for all the cases listed in
 Table~\ref{table:simulation_parameters}, where the legend of line styles and symbols is
 also given.}
\label{fig:mean_velocity}
\end{figure}

The multi-scale character of wall bounded turbulence, in which
$\nu/u_\tau$ is the length scale relevant to the near-wall flow and
$\delta$ applies to the flow away from the wall, is well known. As
first noted by \citet{Millikan1938}, this scaling behavior and
asymptotic matching lead to the logarithmic variation of mean
streamwise velocity with the distance from the wall in an overlap
region between the inner and outer flow. This ``log law'' is given by
\begin{equation}
U^+  =  \frac{1}{\kappa} \log y^+ + B
\label{eq:log_law}
\end{equation}
where $\kappa$ is the \karman constant. In a log layer, the indicator
function
\[
\beta(y^+)= y^+ \frac{\partial U^+}{\partial y^+}
\label{eq:karman_const}
\]
is constant and equal to $1/\kappa$.  Hence, the indicator function,
$\beta$, will have a plateau if there is a logarithmic layer.

\begin{figure}
\centering
\includegraphics[width= \textwidth]{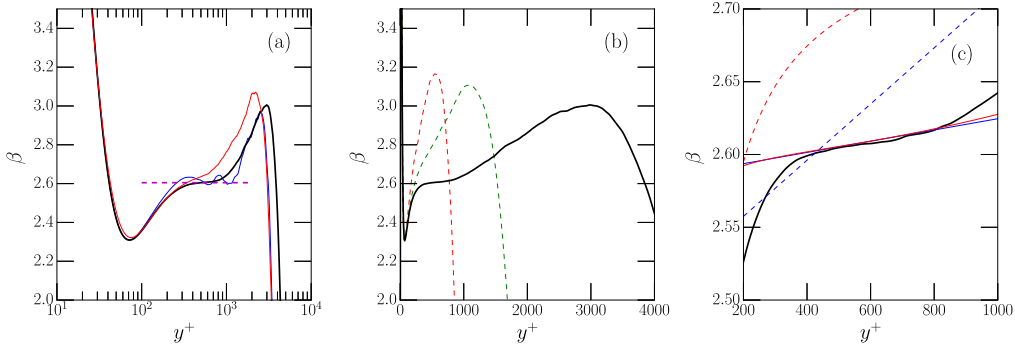} 
\caption{Log-law indicator function $\beta$ for (a) the highest
  Reynolds number simulations, LM5200, LJ4200 and BPO4100, (b)
  simulations at $Re_\tau=5186$, 2003 and 1000 (LM5200, HJ2000 and
  LM1000), and (c) simulation LM5200 along with the
    expressions (\ref{eq:finite_re_correction}, blue) and
    (\ref{eq:Mizuno_correction}, red). In (a), the horizontal dashed
    line is at $\beta=1/\kappa=1/0.384$, and in (c) parameter values
    (see Table~\ref{tab:paramtable}) fit from LM5200 are solid, and from
    \citet{Jimenez2007,Mizuno2011} are dashed.  The linestyle legend for (a) and (b) is
    given in Table~\ref{table:simulation_parameters}.}
\label{fig:indicator}
\end{figure}

The mean streamwise velocity profile is shown in
figure~\ref{fig:mean_velocity} for all the data sets listed in
Table~\ref{table:simulation_parameters}. The profiles from all the
relatively high Reynolds number cases are consistent as expected.
Despite this agreement, the indicator function $\beta$ shows some
disagreement between the three highest Reynolds number simulations, as
shown in figure~\ref{fig:indicator}(a). In the LM5200 case, $\beta$ is
approximately flat between $y^+ = 350$ and $y/\delta=0.16$
($y^+=830$), indicating a log-layer in this region. The LJ4200 case
also appears to be converging toward a plateau in this region, but
there is apparent statistical noise in the profile, which is
understandable given the small domain size. However, the BPO4100
simulation does not have a plateau in $\beta$. There is also a small
discrepancy between BPO4100 and the other two cases from $y^+\approx
30$ to 100. These discrepancies are significantly larger than the
statistical uncertainty in the value of $\beta$ (approximately 0.2\%)
in the current simulations, and presumable in the BPO4100 simulations,
as the averaging time and domain sizes are comparable. 

The values of $\kappa$ and $B$ in (\ref{eq:log_law}) were determined
by fitting the mean velocity data from LM5200 in the region between
$y^+=350$ and $y/\delta=0.16$ to obtain the values of $0.384 \pm
0.004$ and 4.27, respectively, with $R^2 = 0.9999$ where $R^2$ is the
coefficient of determination, which is one for a perfect fit. The
value of $\kappa$ agrees with the value computed by
\citet{Lozano-Duran2014}, but shows a slight discrepancy with the
values measured by \citet{Nagib2008} and \citet{Montythesis}, which
are $\kappa=0.37$ and 0.39, respectively.  The value of $\kappa$
obtained here is remarkably similar to that reported by
\citet{Osterlund2000}, $\kappa=0.38$, and, \citet{Nagib2008}, $\kappa
= 0.384$, in the zero pressure gradient boundary layer. However, the
value of $\kappa$ reported here is smaller than $\kappa = 0.40$
measured by \citet{Bailey2014} in pipe flow. Note that choosing the
range for this curve fit is somewhat arbitrary, since the indicator
function is not exactly flat (figure~\ref{fig:indicator}c).

From an asymptotic analysis perspective, the log-law relation
(\ref{eq:log_law}) is the lowest order truncation of a matched
asymptotic expansion in $1/Re_\tau$ \citep{Afzal1973,Jimenez2007}. Several higher order
representations of the mean velocity in the overlap region have been
evaluated based on experimental and DNS data in boundary layers,
channels and pipes \citep{Buschmann2003,Jimenez2007,Mizuno2011}. Here
we consider two such applications to channels in the context
of the LM5200 data. 

\citet{Jimenez2007}  considered a higher order truncation
in which $\beta$ has the form:
\begin{equation}
\beta = y^+ \frac{\partial U^+}{\partial y^+}  =  \left(\frac{1}{\kappa_\infty} +
\frac{\alpha_1}{Re_\tau}\right) + \alpha_2\frac{y^+}{Re_\tau}
\label{eq:finite_re_correction}
\end{equation}
This formulation essentially allows for an $Re_\tau$ dependence of
$\kappa=(1/\kappa_\infty+\alpha_1/Re_\tau)^{-1}$, and introduces a
linear dependence on $y/\delta=y^+/Re_\tau$. Based on data from a
simulation at $Re_\tau\approx1000$ by \citet{DelLamo2004} (similar to
LM1000) and from HJ2000, (see figure~\ref{fig:indicator}b), they
determined the parameter values shown in
table~\ref{tab:paramtable}. Further, this form and these values were
found to be consistent with experimental measurements by \citet{Christensen2001} at
Reynolds numbers up to $Re_\tau=2433$.

\citet{Mizuno2011}
considered a different higher-order asymptotic truncation, for which
$\beta$ is given by:
\begin{equation}
\beta = y^+ \frac{\partial U^+}{\partial y^+} = \frac{y^+}{\kappa(y^+-a_1)} + a_2\frac{y^{+2}}{Re_\tau^2}.
\label{eq:Mizuno_correction}
\end{equation}
The second term was motivated by the form of a wake
model that is quadratic for small $y/\delta$, where the coefficient is related to the
wake parameter $\Pi$ by $a_2=(12\Pi-2)/\kappa$. The term $a_1$ is an
offset (virtual origin) which accounts for the presence of the
viscous layer \citep{Wosnik2000}. To first order in $a_1/y^+$, the offset is
equivalent to including an additive $a_1/\kappa y^+$ term, which is
expected from the matched asymptotics.  They fit the inverse of the
mean velocity derivative to $y^+/\beta$ from (\ref{eq:Mizuno_correction}) using the experimental and DNS data mentioned
above and the experiments of \citet{Montythesis}, with up to
$Re_\tau=3945$, to determine a
Reynolds number dependent value of the parameters. When evaluated for
$Re_\tau=5186$, these yield the parameter values shown in
table~\ref{tab:paramtable}.

\begin{table}
  \begin{center}
    \def~{\hphantom{0}}
%    \resizebox{\textwidth}{!}{
      \begin{tabular}{lccp{1cm}lcc}
        \multicolumn{3}{c}{Equation (\ref{eq:finite_re_correction})}&
        &\multicolumn{3}{c}{Equation (\ref{eq:Mizuno_correction})} \\
        \noalign{\vskip 2pt} 
        \cline{1-3} \cline{5-7}
        \noalign{\vskip 2pt} 
        \noalign{\vskip 2pt} 
        &\citet{Jimenez2007}&LM5200&{     }& &\citet{Mizuno2011}&LM5200\\
        \noalign{\vskip 2pt} 
        $\kappa_\infty$  &  ~~0.402~ &  & & $\Pi$    & ~~0.179 & ~0.189 \\
        $\alpha_1$       &  150.~~~~ &  & & $a_1$    & -12.4~~ & -1.0~~ \\
        $\alpha_2$       &  ~~1.0~~~ & 0.2~~~ & & $a_2$    & ~~0.394 & ~0.7~~ \\
        $\kappa$         &  ~~0.3970 & 0.3867 & & $\kappa$ & ~~0.363 & ~0.384 \\
%        \noalign{\vskip 2pt} 
%        \noalign{\hrule}
%        \noalign{\vskip 2pt}
%        \noalign{\vskip 2pt} 
      \end{tabular}
%    }
  \caption{Values of parameters in (\ref{eq:finite_re_correction}) and (\ref{eq:Mizuno_correction})
    appropriate for $Re_\tau=5186$ as determined by \citet{Jimenez2007,Mizuno2011} and
    from the LM5200 data.}
  \label{tab:paramtable}
  \end{center}
\end{table}

These two higher-order truncations have also been fit to the LM5200
data to obtain values shown in table~\ref{tab:paramtable}, which are
significantly different from the previously determined values. The
expressions for $\beta$ from (\ref{eq:finite_re_correction})  and (\ref{eq:Mizuno_correction}) are plotted in
figure~\ref{fig:indicator}(c) with both sets of parameters. It is
clear from this figure that the parameter values obtained by
\citet{Jimenez2007} and \citet{Mizuno2011} do not fit the LM5200 data, but the
parameters fit to the LM5200 data in the log region match the data
equally well for both truncation forms. The reason for this
disagreement with the parameters from \citet{Jimenez2007} is clear,
since the Reynolds numbers used in that study were not high enough to
exhibit the logarithmic region observed in LM5200, since
$y/\delta=0.16$ is at $y^+=320$ when $Re_\tau=2000$. They appear to
have been fitting (\ref{eq:finite_re_correction}) to the outer-layer profile, and indeed
in LM5200 there is a region $0.16<y/\delta<0.45$ in which $\beta$ is
approximately linear in $y^+/Re_\tau$, with slope of one
(figure~\ref{fig:indicator}b), in agreement with $\alpha_2=1.0$.

In contrast, the data used by \citet{Mizuno2011} included a channel Reynolds
number as high as $Re_\tau=3945$, which should have exhibited a
short nearly constant $\beta$ plateau as observed in LM5200. This
would have been qualitatively different from the lower Reynolds number
cases. This was not reported in in that paper. But, at this Reynolds
number, the plotted values of $\kappa$ and $a_1$ (figure~5 in
\citet{Mizuno2011}) are significantly larger than the parameters for lower
Reynolds number. Further, the values for all three parameters obtained from
the LM5200 data are within the indicated uncertainties of the
parameters from the $Re_\tau=3945$ experimental data. Perhaps the
reported Reynolds number dependence of the parameters did not reflect
a qualitative change at the highest Reynolds number, because the fit
was dominated by the more numerous lower Reynolds number cases.
  
The apparent extent of the overlap region  
in the
LM5200 case is not sufficient to distinguish between the two
asymptotic truncations, (\ref{eq:finite_re_correction}) and (\ref{eq:Mizuno_correction}). Because $a_1$
is so small, the primary distinction is in the lowest non-zero
exponent on $y^+/Re_\tau$. This may be of some importance because it
determines the way the high Reynolds number asymptote of constant
$\beta$ is approached. Unfortunately, this cannot be determined using
the available data.

%% file: velocity_fluctuation.tex
\begin{figure}%[ht]
\centering
\includegraphics[width=.85\textwidth]{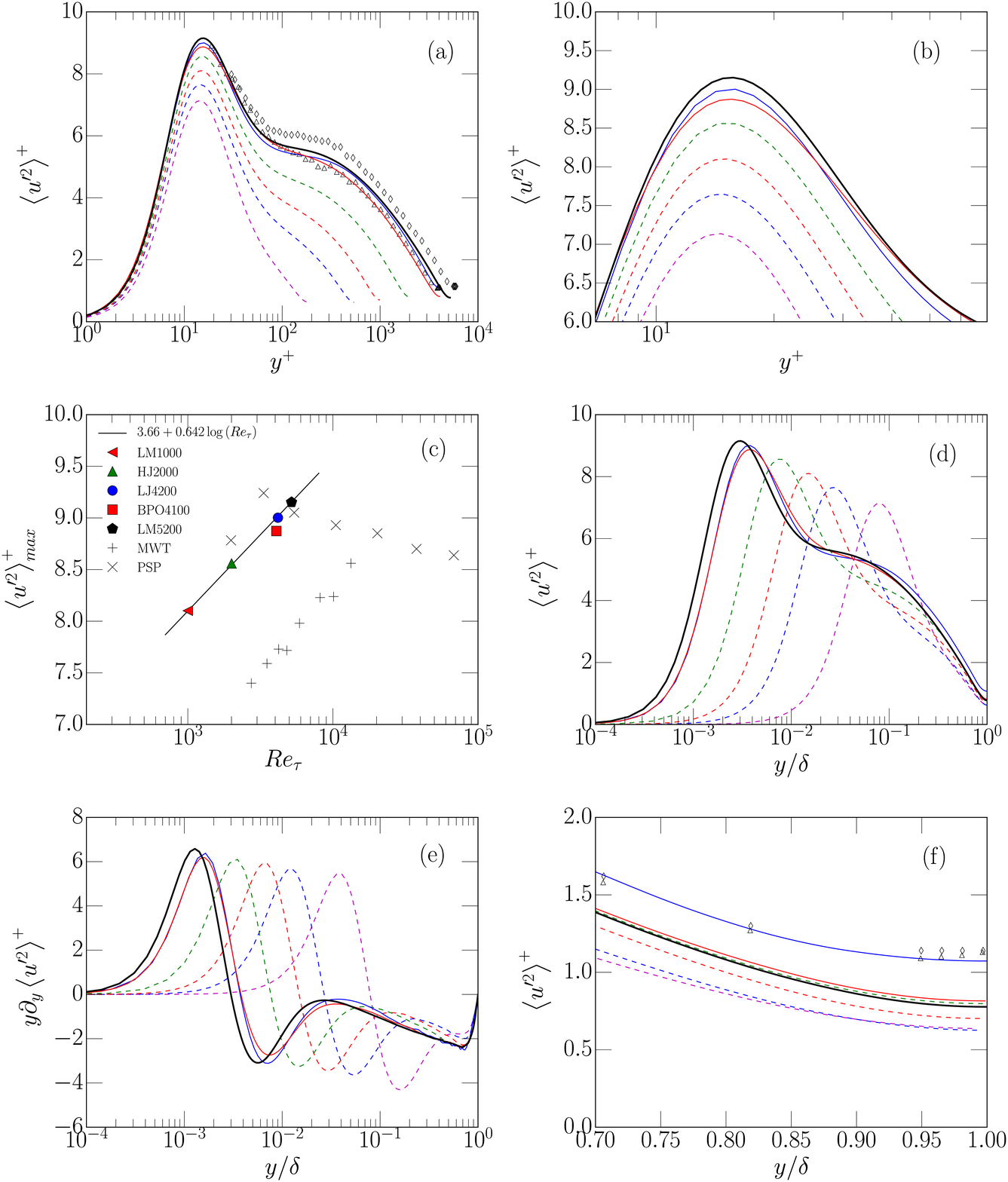}
\caption{Variance of $u$:
(a) as a function of $y^+$;
(b) zoom of (a) near the peak;
(c) dependence of maximum on $Re_\tau$; Solid line, relation;
(\ref{eq:uupeak}); MWT, boundary layer in the Melbourne Wind Tunnel \citep{Hutchins2009,Kulandaivelu2011}; PSP,
Princeton Superpipe \citet{Hultmark2010,Hultmark2012};
(d) as a function of $y/\delta$;
(e) test function for Townsend's prediction;
and, (f) zoom of (d) near the center of channel. The linestyle and symbol
legend is given in Table~\ref{table:simulation_parameters}.}
\label{fig:uu}
\end{figure}
\begin{figure}
\centering
\includegraphics[width=\textwidth]{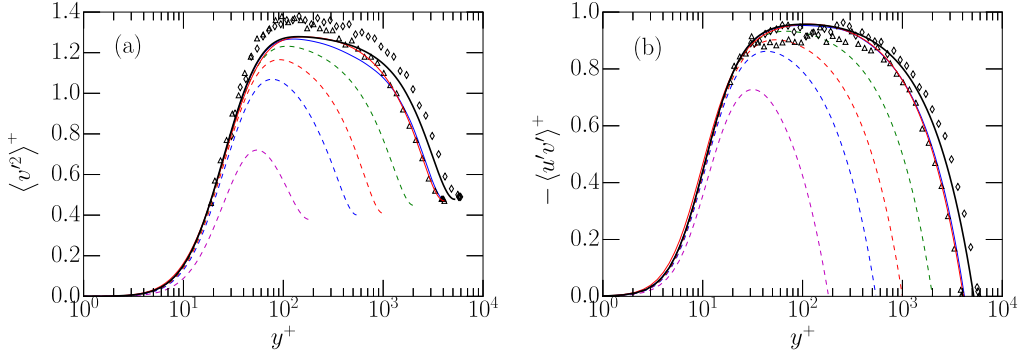}
\caption{Variance of $v$ (a) and covariance of $u$ and $v$ (b).
The linestyle and symbol legend is given in
Table~\ref{table:simulation_parameters}}.
\label{fig:vv_uv}
\end{figure}
\begin{figure}
\centering
\includegraphics[width=\textwidth]{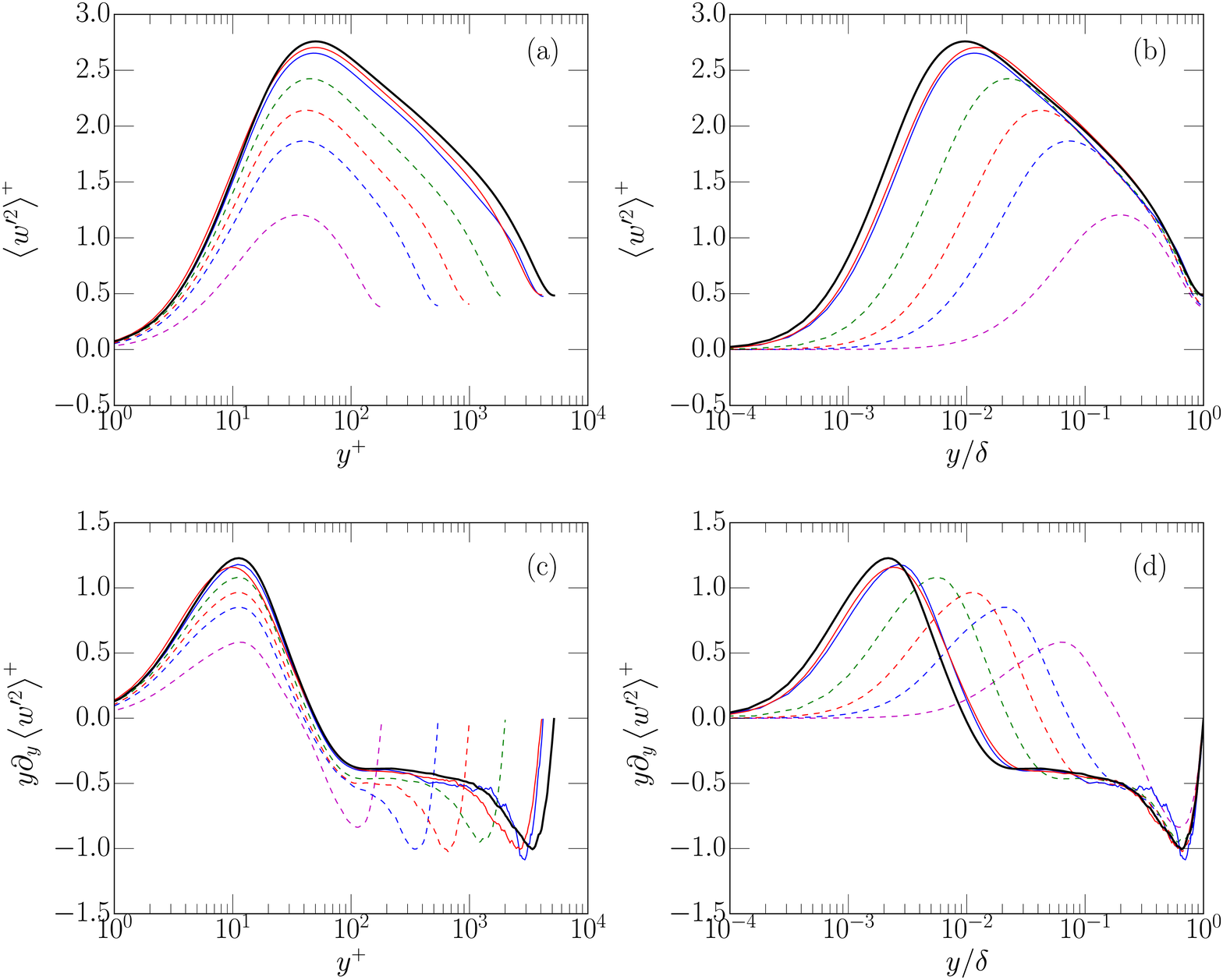}
\caption{Variance of $w$
(a) as a function of $y^+$
(b) as a function of $y/\delta$, and
test function for Townsend's prediction
(c) as a function of $y^+$ 
(d) as a function of $y/\delta$. The linestyle and symbol legend is
given in Table~\ref{table:simulation_parameters}.
}
\label{fig:ww}
\end{figure}

The non-zero components of the Reynolds stress tensor (the velocity
component variances and co-variance) from the simulations and
experiments are shown in Figs.~\ref{fig:uu}--\ref{fig:ww}. These
figures show that there are some subtle inconsistencies among the
three highest Reynolds number simulations (solid lines in the figures)
and the experimental data.

First, the two cases LJ4200 and BPO4100 are nearly identical in
Reynolds number, but all three velocity variances are different
between the two cases. The peak of the streamwise variance
(Fig.~\ref{fig:uu}b) is about 1.4\% larger in LJ4200 than in
BPO4100. The peak varies with Reynolds number, as can be seen in the
figure, but this variation is logarithmic, and is too weak to explain
the difference. Using the simulations LM1000, HJ2000 and LM5200, the
dependence of the peak in $\langle u^{\prime 2}\rangle^+$ on $Re_\tau$
was fit to get
\begin{equation}
\langle u^{\prime 2} \rangle^+_{max} = 3.66+0.642\log\left(Re_\tau\right)
\label{eq:uupeak}
\end{equation}
with $R^2=0.9995$. This agrees well with the relationship, $\langle
u^{\prime 2} \rangle^+_{max} = 3.63+0.65\log\left(Re_\tau\right)$,
suggested by \citet{Lozano-Duran2014}. The relationship
(\ref{eq:uupeak}) is plotted in Fig~\ref{fig:uu}c along with the
values of the actual peaks, including those for LJ4200 and BPO4100. It
is clear from this plot that the peak in BPO4100 is smaller than this
relationship implies for $Re_\tau=4100$, while the peak from LJ4200 is
consistent. 

Also shown in Fig~\ref{fig:uu}(c) are experimental data
for pipe flow by \citet{Hultmark2010,Hultmark2012}, which indicate
that the inner-peak value of $\left< u^{\prime 2}\right>^+$ does not
continue to increase with $Re$ for $Re_\tau>5000$. However,
experimental data from boundary layers
by \citep{Hutchins2009,Kulandaivelu2011} do not show such a growth
saturation.  The Reynolds numbers of the current simulations are
unfortunately not high enough to determine whether the growth of
$\left< u^{\prime 2}\right>^+$ will saturate in channel flows. Also
remarkable in Fig~\ref{fig:uu}(c) is how much lower the peak values
are in the boundary layer data. This is likely due to the resolution
of the measurements \citep{Hutchins2009}.
\begin{comment}
\todo{need a reference for these, private
communication with Ivan if necessary, MK:Done}
\end{comment}

In other $y^+$ intervals ($y^+<8$ and $25<y^+<200$), the variance in
the lower Reynolds number case (BPO4100) is actually greater than the
variance in the higher Reynolds number flow (LJ4200). This appears to
be inconsistent with the Reynolds number trends among the other cases,
for which $\langle u^{\prime 2}\rangle^+$ is monotonically increasing
with Reynolds number at constant $y^+$.  Another inconsistency is
apparent when the streamwise velocity variance is examined as a
function of $y/\delta$ (Fig.~\ref{fig:uu}d,f). Near the centerline
($y/\delta=1$), the variance from LJ4200 is significantly larger than
for both BPO4100 and LM5200, while the other simulations indicate that
variance should be increasing slowly with Reynolds number. It appears
that far from the wall, LJ4200 is affected by its relatively small
domain size, as might be expected. Finally, the experimental data with
reported uncertainty, $\pm$ 2\%, from SF4000 in Fig~\ref{fig:uu}a is
inconsistent with LJ4200 and BPO4100 in the region near
the wall ($y^+<300$ say).  The reason for this discrepancy is not
clear, but it may be due to the difficulty of measuring velocity fluctuations 
near the wall. Far from the wall, the experimental data
for these quantities is consistent with the simulations.  Measurements
are not available at small enough $y^+$ to compare peak values of
$\langle u^{\prime2}\rangle^+$.

Similar inconsistencies are present among the high Reynolds number
simulation cases in the wall-normal and spanwise velocity variances.
Around the peaks of both $\langle v^{\prime 2}\rangle^+$ and $\langle
w^{\prime 2}\rangle^+$, BPO4100 exceeds values from LJ4200, despite
its somewhat lower Reynolds number (Fig.~\ref{fig:vv_uv}a and
Fig.~\ref{fig:ww}a), while near the center, these two cases are in
agreement. Only the Reynolds shear stress $\langle u'v'\rangle^+$ is
in agreement in these cases across all $y$
(Fig~\ref{fig:vv_uv}b). Similar to $\langle u^{\prime 2} \rangle ^+$,
the experimental data from SF4000 for $\langle v^{\prime 2}\rangle^+$
and $\langle u'v'\rangle^+$ with uncertainties, $\pm$ 3\% and $\pm$
5\%, respectively, are inconsistent with both simulations in the region
near the wall, where the experimental data appears to be quite noisy.
Possible reasons for the minor inconsistencies noted among the DNS are
discussed in \S\ref{sec:resolution}.
\begin{comment}
\todo{MK: Should we remove this paragraph? RDM: OK, then added sentence how above, OK?}
\end{comment}
\begin{comment}
Overall then, the comparisons discussed above suggest that the BPO4100
case is inconsistent with other simulations near the wall and the
LJ4200 case is inconsistent far from the wall, presumably due to its
small domain size. See \S\ref{sec:resolution} for a more detailed
discussion of plausible reasons for these inconsistencies. There are no
obvious inconsistencies with the LM5200 simulation.
\end{comment}

There are several anticipated high-Reynolds-number features to be
examined in this data.  In particular, similar to the log law,
Townsend's attached eddy hypothesis \citep{Townsend1976} implies that
in the high Reynolds number limit, there is an interval in $y$ in
which the Reynolds stress components satisfy
\begin{subeqnarray}
\langle u^{\prime 2}\rangle^+ & = & A_1 - B_1 \log\left(y/\delta\right) \\[3pt]
\langle v^{\prime 2}\rangle^+ & = & A_2 \\[3pt]
\langle w^{\prime 2}\rangle^+ & = & A_3 - B_3 \log\left(y/\delta\right) \\[3pt]
\langle u'v' \rangle^+ & = & -1
\label{eq:townsend_prediction}
\end{subeqnarray}
Consistent with these relations, both $\langle v^{\prime2}\rangle^+$
and $-\langle u'v'\rangle^+$ are developing a flat region as Reynolds
number increases (Fig~\ref{fig:vv_uv}), though the maximum Reynolds
shear stress is well bellow one (0.96). Because the total stress
\begin{equation}
\tau_{\rm tot}=\nu\frac{\partial U}{\partial y}-\langle u'v'\rangle
\label{eq:tautotal}
\end{equation}
is known analytically ($\tau^+_{\rm tot}= 1-y/\delta$) from the mean
momentum equation, and because $\beta$ varies little over a broad
range of $y$ ($\beta=2.6\pm 0.4\approx 1/\kappa$ for
$30<y^+<0.75Re_\tau$ (see Fig.~\ref{fig:indicator}), the variation
with Reynolds number of the Reynolds shear stress near its maximum can
be deduced easily:
\begin{equation}
\tau_{RS}^+ = -\langle u'v'\rangle^+\approx 1-\frac{y^+}{Re_\tau}-\frac{1}{\kappa y^+}\quad\mbox{for}\quad 30<y^+<0.75Re_\tau
\label{eq:RSmodel}
\end{equation}
From this it is clear that the maximum Reynolds shear stress is given
by $\tau_{RS{\rm max}}\approx1-2/\sqrt{\kappa Re_\tau}$, and that this
maximum occurs at $y^+\approx \sqrt{Re_\tau/\kappa}$ as noted
by \citet{Afzal1982}, \citet{Morrison2004}, \citet{Panton2007}
and \citet{Sillero2013}.  For the conditions of LM5200
($Re_\tau=5186$, $\kappa=0.384$), these estimates yield $\tau_{RS{\rm
max}}\approx 0.955$ occurring at $y^+\approx 116$, in good agreement
with the simulations. Further, the error in satisfying $\tau_{RS}=1$
is less than $\epsilon$ for a range of $y^+$ that increases in size
like $\epsilon Re_\tau$ for large $Re_\tau$. Precisely:
\begin{equation}
1-\tau_{RS}<\epsilon\quad\mbox{provided}\quad
\left|y^+-\frac{\epsilon Re_\tau}{2}\right|<\frac{\epsilon Re_\tau}{2}\sqrt{1-\frac{4}{\epsilon^2 Re_\tau\kappa}}.
\end{equation}
Thus, for $\tau_{RS}$ to be within 5\% of one over a decade of
variation of $y^+$ would require more than twice the Reynolds number
of LM5200, and for it to be within 1\% at its peak requires 20 times
greater Reynolds number.

According to (\ref{eq:townsend_prediction}), both the variance of $u$
and the variance of $w$ would have a logarithmic variation over some
region of $y$. In Fig.~\ref{fig:ww}, it appears that there is such a
logarithmic variation, even at Reynolds numbers as low as
$Re_\tau=1000$. The indicator function $y\partial_y\langle w^{\prime
2}\rangle$ is approximately flat from $y^+\approx 100$ to $y^+\approx
200$ (Fig.~\ref{fig:ww}c). The corresponding curve fit is $\langle w^{\prime2} \rangle ^+ = 1.08
- 0.387 \log (y/\delta)$, which is somewhat different from the fit $\langle w^{\prime2} \rangle ^+ =
0.8 - 0.45 \log (y/\delta)$ obtained by \citet{Sillero2013} in a
boundary layer DNS. On the other hand, there is no apparent
logarithmic region in the $\langle u^{\prime 2}\rangle^+$
profiles. The indicator function $y\partial_y\langle u^{\prime
2}\rangle^+$ (Fig.~\ref{fig:uu}e) is not flat anywhere in the
domain. There is however a region ($200<y^+<0.6Re_\tau$), where the
dependence of the indicator function on $y$ is linear with relatively
small slope, and the slope may be decreasing with Reynolds number,
though extremely slowly. In
contrast, \citet{Hultmark2012,Hultmark2013} observed a logarithmic region in
$\langle u^{\prime 2}\rangle^+$ over the range
$800<y^+<0.15Re_\tau$
in pipe flow with
$Re_\tau>2\times10^4$. The LM5200 simulation is not at high enough Reynolds
number to exhibit such a region if it occurs over the same range in $y$,
since $0.15Re_\tau<800$ at $Re_\tau=5186$.

\begin{figure}
\centering
	\includegraphics[width=.7\textwidth]{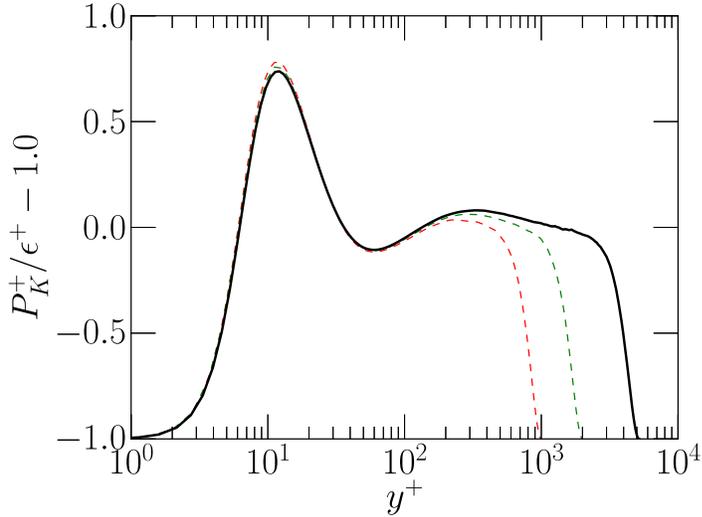}
\caption{Balance of production and dissipation of turbulent kinetic energy}
\label{fig:budget}
\end{figure}

As mentioned in \S\ref{sec:method}, the terms in the Reynolds stress
transport equations are not reported here, though the data are
available at {\tt http://turbulence.ices.utexas.edu}.  Of interest
here, however, is the transport equation for the turbulent kinetic
energy $K=\frac{1}{2}\langle u'_iu'_i \rangle$. \citet{Hinze1975}
argues that at sufficiently high Reynolds number, there is an
intermediate region between inner and outer layers where the transport
terms in the kinetic energy equation are small compared to production
so that in this region
\begin{equation}
P_K\approx {\epsilon},
\label{balance}
\end{equation}
where $P_K$ is the production of kinetic energy and $\epsilon$ is the
dissipation.  In the formulation and analysis of turbulence models,
(\ref{balance}) is often assumed to hold in an overlap region between
inner and out layers \citep{Durbin2010}.  The relative error in the
balance of production and dissipation ($P_K/\epsilon-1$) for LM1000,
HJ2000 and LM5200, is shown in figure~\ref{fig:budget}. In all three
cases, there is a region $y^+>30$ and $y/\delta<0.6$ in which the
mismatch between production and dissipation is of order 10\% or less,
but there is no indication that the magnitude of this mismatch is
decreasing with Reynolds number. Indeed, there appears to be a stable
structure with a local minimum in $P_K/\epsilon$ around $y^+=60$ and a
local maximum around $y^+=300$ followed by a gradual decline toward
$P_K/\epsilon=1$ with increasing $y$. Presumably this decline will
become more gradual with increasing Reynolds number as $y/\delta=0.6$
increases in $y^+$.

%% file: resolution.tex
In \S\ref{sec:mean} and \S\ref{sec:fluctuations}, it was shown that
the LM5200 simulation has differences from the two somewhat lower
Reynolds number DNS LJ4200 and BPO4100, with discrepancies among all
three simulations larger than expected given the Reynolds number
differences and likely statistical errors. In the case of LJ4200, the
reason for the differences with LM5200 is most likely the small domain
size of LJ4200 (four and three times smaller in the streamwise and
spanwise directions respectively). Indeed \citet{Lozano-Duran2014}
investigated the effects of the small domain size at $Re_\tau \approx
950$ by comparing to DNS in a domain consistent with the simulations
reported here. Consistent with observations
in \S\ref{sec:fluctuations}, they observed discrepancies in the
velocity variances for $y/\delta< 0.25$.

The differences between BPO4100 and both LJ4200 and LM5200 occur
primarily in the near-wall region, as shown
in \S\ref{sec:fluctuations}. It seems likely that these differences
are due to numerical resolution limitations of BPO4100. As indicated
in Table~\ref{table:simulation_parameters}, the grid spacing in
BPO4100 is about the same as LM5200 in the spanwise direction and
about 25\% finer in the streamwise direction. However, in BPO4100 a
second order finite difference approximation is used rather than the
spectral method used in the simulations reported here. The effective
resolution of such low-order schemes is significantly less at the same
grid spacing. One way to see this is to consider the error incurred
when differentiating a sine function of different wavenumbers, which
is given by $\epsilon(k)=1-\sin(k\Delta)/k\Delta$. If one insists on
limiting error to no more than 10\% (for example), then the wavenumber
needs to be limited to one fourth of the highest wavenumber that can
be represented on the grid.  Thus, at this level of error, the finite
difference approximation of the derivative has four times coarser
effective resolution than a spectral method on the same grid. 

Of course, there is much more to solving the Navier-Stokes equations
than representing the derivative and an appropriate error limit for
the derivative is not clear. None-the-less, in DNS, a common rule of
thumb is that second order finite difference has between two to four
times coarser effective resolution than a Fourier spectral method on
the same grid. In a study of the effects of resolution in DNS of low
Reynolds number ($Re_\tau=180$) channel flow, \citet{Oliver2014} found
that with the Fourier/B-spline numerical representation used here,
coarsening the resolution by a factor of two relative to the nominal
resolution (nominal is comparable in wall units to that used for
LM5200) results in changes of several percent in the
velocity variances near the wall. Further, \citet{Vreman2014}
investigated differences between DNS using a Fourier-Chebyshev
spectral method and DNS using a fourth order staggered finite
difference method, which is higher order and higher resolution than
the method used for BPO4100. They reported a 1\% lower peak rms $u'$
for the fourth-order method. These results indicate that the lower
effective resolution in BPO4100 relative to LM5200 is a plausible
cause for the minor inconsistencies between these two simulations.
To determine this definitively would require redoing the BPO4100
simulation with twice the resolution in each direction, or more, which
is out of scope for the current study.

%% file: energy_spectra.tex
\begin{figure}
\begin{center}
	\includegraphics[width=\textwidth]{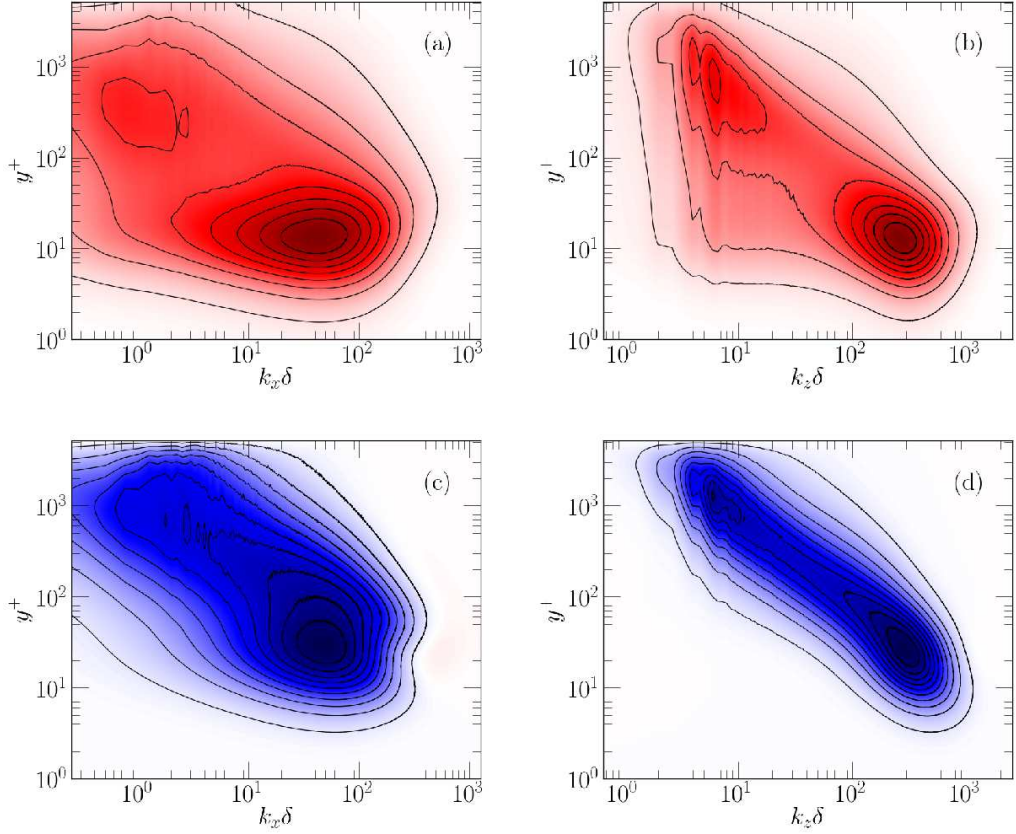}
\end{center}
\caption{Wavenumber-premultiplied energy spectral density, LM5200 
  (a) $k_x E_{uu}/u_\tau ^2$ 
  (b) $k_z E_{uu}/u_\tau ^2$
  (c) -$k_x E_{uv}/u_\tau ^2$ 
  (d) -$k_z E_{uv}/u_\tau ^2$}
\label{fig:kEuu_5200}
\end{figure}

One of the properties of high-Reynolds-number wall-bounded turbulence
is the separation of scales between the near-wall and outer-layer
turbulence. For the LM5200 case, this separation of scales can be seen
in the one-dimensional velocity spectra. For example, the
premultiplied spectral energy density of the streamwise velocity
fluctuations is shown as a function of $y$ in
figure~\ref{fig:kEuu_5200}.  The pre-multiplied spectrum $kE(k,y^+)$ is
the energy density per $\log k$ and so, in the logarithmic wavenumber
scale used in figure~\ref{fig:kEuu_5200}, it indicates the scales at
which the energy resides. The energy spectral density of $u$ in the
streamwise direction has two distinct peaks, one at $k_x \delta=40$,
$y^+ =13$ and one at $k_x \delta = 1$, $y^+ = 400$. To our knowledge,
such distinct peaks have only previously been observed in high
Reynolds number experimental
data \citep{Hutchins2007,Monty2009a,Marusic2010,Marusic2010b}. An
even more vivid double peak is visible in the spanwise spectrum of
$u$ (figure~\ref{fig:kEuu_5200}b), with peaks at $k_z \delta=250$,
$y^+ =13$ and at $k_z \delta = 6$, $y^+ = 1000$.

Similarly there are weak double peaks in the cospectrum of $uv$ as
shown in figure~\ref{fig:kEuu_5200}c,d. In the LM5200 case, distinct
inner and outer peaks were not observed in the spectral density of
$v^2$ and $w^2$ (not shown). It thus appears that it is primarily the
streamwise velocity fluctuations that are exhibiting inner/outer scale
separation at this Reynolds number.

\begin{figure}
\begin{center}
	\includegraphics[width=\textwidth]{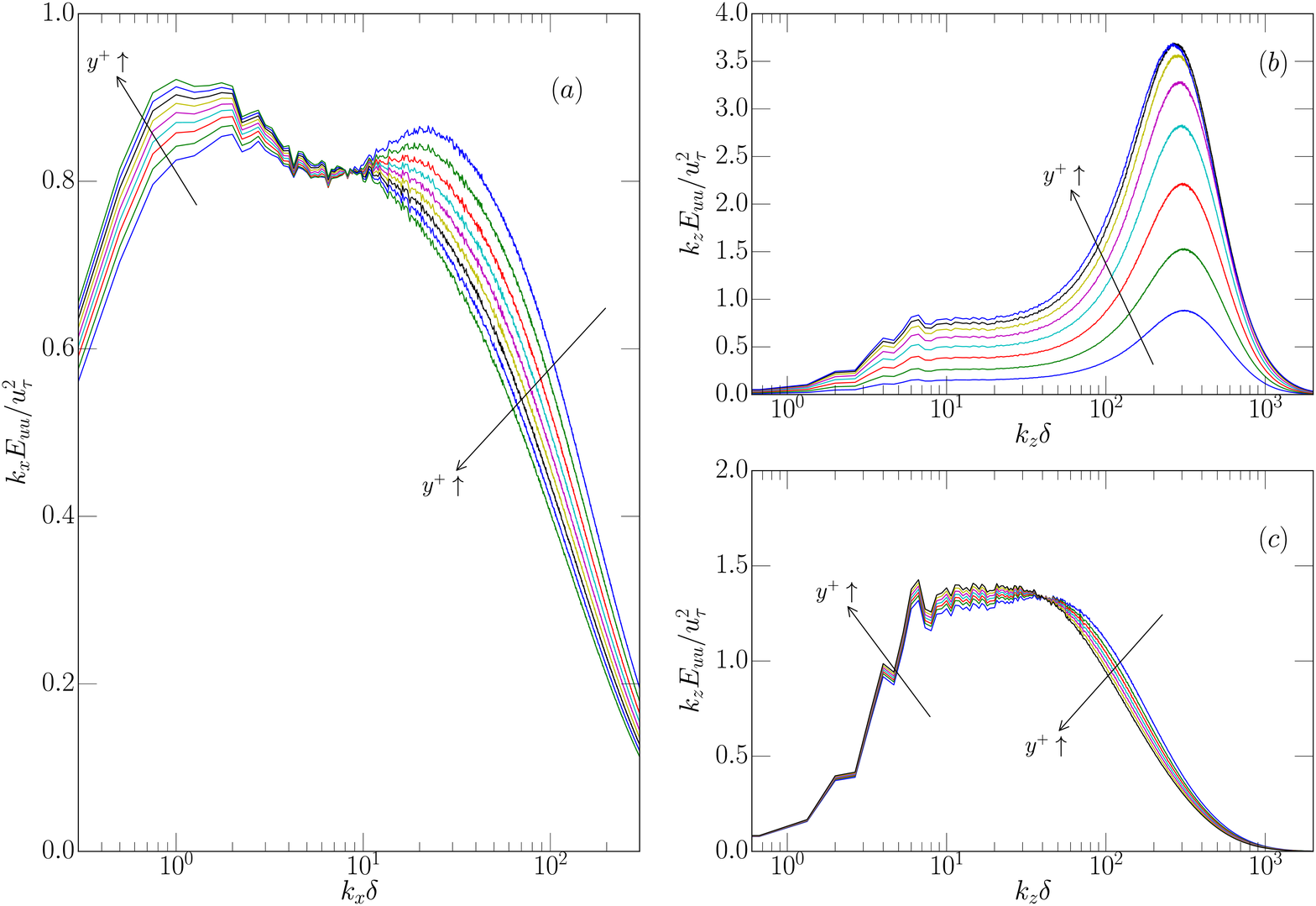} 
\end{center}
\caption{$k^{-1}$ region (a) $k_x  E_{uu}/u_\tau^2$ at $y^+ =  90\sim170$ 
(b) $k_z  E_{uu}/u_\tau^2$ at $y^+ =  3 \sim 14$
(c) $k_z  E_{uu}/u_\tau^2$ at $y^+ =  111 \sim 141$}
\label{fig:k1_region}
\end{figure}

As mentioned earlier, the scaling analysis of \citet{Perry1986}
suggests that at high Reynolds number the energy spectral density of
the streamwise velocity fluctuations varies as $k_x^{-1}$ in the
overlap region where both inner and out scaling is valid. However,
until now, the $k_x^{-1}$ region has been elusive in simulations,
presumably because the Reynolds numbers have not been high enough.  It
is only in high Reynolds number experiments
\citep{Nickels2005,Nickels2007,Rosenberg2013} that a
$k_x^{-1}$ regions has previously been observed in the streamwise
velocity spectrum. In the LM5200 simulation, the premultiplied energy
spectral density does indeed exhibit a plateau in the region $90\le
y^+\le 170$ and $6\le k_x \delta\le 10$ (figure~\ref{fig:k1_region}
(a)).  Further, the magnitude of the premultiplied spectrum at the
plateau agrees with the value observed experimentally, about 0.8 in
wall units, in boundary layers \citep{Nickels2005} and in
pipe flow \citep{Rosenberg2013}. 
However, the plateau in
the experimental premultiplied spectrum in the pipe flow
of \citet{Rosenberg2013} is only observed for
$Re_\tau \le 3300$.
From the current results, we cannot determine whether
 the occurrence or extent of the plateau region might change
 at higher Reynolds number in channels.

A similar scaling analysis also suggests that there
should be a plateau in the premultiplied spanwise spectrum, and indeed
an even broader plateau appears for $5\le k_z \delta\le 30$, as
observed previously by \citet{Hoyas2008} and \citet{Sillero2013}.
Interestingly, the plateau also occurs in the viscous sub-layer region
(figure~\ref{fig:k1_region}~(b)). In the viscous sub-layer the
streamwise velocity fluctuations are dominated by the well-known
streaky structures, with a characteristic spacing of $\Delta
z^+\approx 100$. This is evidenced by the large high-wavenumber peaks
in the spanwise spectra around $k_z/\delta\approx 300$ ($k_z^+\approx
2\pi/\Delta z^+$). At much larger scales (much lower wavenumbers), the
viscous layer is driven by the larger-scale turbulence further from
the wall. The plateau in the spanwise premultiplied spectrum in the viscous layer is
thus a reflection of the spectral plateau further from the wall
(figure~\ref{fig:k1_region}~(c)).

Another experimentally observed feature of the streamwise
premultiplied spectrum is the presence of two local maxima, with one
peak occurring on either side of the
plateau \citep{Guala2006,Kunkel2006}. However, this bi-modal feature
was called into question when it was noted that the low-wavenumber
peak, which occurs at $k_x\delta$ of order one, is at low enough
wavenumber to be affected by the use of Taylor's hypothesis to infer
the spatial spectrum from the measured temporal
spectrum \citep{DelAlamo2009,Moin2009}. Indeed, in the HJ2000
simulation, it was found that the streamwise spatial spectrum did not
display a low-wavenumber peak, while a spectrum determined from a time
series did \citep{DelAlamo2009}.  In the LM5200 case, this bi-modal
feature does occur, as can be seen in
figure~\ref{fig:k1_region}~(a). However, it is not clear whether this
is a general high Reynolds number feature, or whether it is
characteristic of intermediate Reynolds numbers. From
figure~\ref{fig:kEuu_5200}, it is clear that the high and low
wavenumber peaks in $k_x E_{uu}(k_x ,y^+)$ at around $y^+=100$ are
simply the upper and lower edges of the near-wall and outer peaks in
$k_x E_{uu}(k_x ,y^+)$, respectively. As the Reynolds number
increases, the inner peak will move to larger $k_x $ and the outer
peak to larger $y^+$. As the outer peak moves to larger $y^+$, it
seems likely that the inner and outer peaks will not overlap at all in
$y$, resulting in no bi-modal pre-multiplied spectrum at any $y$. For
example, the peaks are better separated in the premultiplied
spanwise spectrum, and there are no bi-modal $k_z$
pre-multiplied spectra (figure~\ref{fig:k1_region} (b)).

%% file: conclusions.tex
The direct numerical simulation of turbulent channel flow at
$Re_\tau=5186$ that is reported here has been shown to be a reliable
source of data on high Reynolds number wall-bounded turbulence, and a
wealth of statistical data from this flow is available on line at {\tt
http://turbulence.ices.utexas.edu}.  In particular: resolution is
consistent with or better than accepted standards for wall-bounded
turbulence DNS; statistical uncertainties are generally small (of
order 1\% or less), statistical data is consistent with Reynolds
number trends in DNS at $Re_\tau\leq 2000$, and with experimental
data, within reasonable tolerance; and, finally, the flow is
statistically stationary to high accuracy. Further, as recapped below,
the simulation exhibits many characteristics of high Reynolds number
wall-bounded turbulence, making this simulation a good resource for
the study of such high Reynolds number flows. Several conclusions can
thus be drawn about high Reynolds number turbulent channel flow as
discussed below.

At high Reynolds number, it is expected that a region of logarithmic
variation will exist in the mean velocity. The current simulation
exhibits an unambiguous logarithmic region with \karman constant
$\kappa=0.384 \pm 0.004$. This is the first such unambiguous simulated
logarithmic profile that the authors are aware of. At high Reynolds
number, it is also expected \citep{Townsend1976} that the variance of
the velocity fluctuations will have a region of logarithmic
variation. This was found to be true for the spanwise fluctuations,
but not the streamwise fluctuations. Indeed, the streamwise variance
shows no sign of converging towards a logarithmic variation with
increasing Reynolds number. Finally, it is often assumed that at high
Reynolds number the production of turbulent kinetic energy will
locally balance its dissipation over some overlap region between inner
and outer layers. However, this was observed only approximately in the
current simulation, with 10\% accuracy and with no sign that that this
miss-match is declining with Reynolds number. This balance of
production and dissipation thus appears to be just an imperfect
approximation.

One of the most important characteristics of high Reynolds number
wall-bounded turbulence is a distinct separation in scale between
turbulence near the wall and far from the wall. This was observed in
the streamwise and spanwise premultiplied spectral density of the
streamwise velocity fluctuations and in the co-spectrum of streamwise
and wall-normal fluctuations. 
Also, a short $k^{-1}$ region is
observed in the streamwise spectra and there is a wider region
in the
spanwise spectra, as predicted from scaling
analysis \citep{Perry1986}, and as observed experimentally. Finally,
the bi-modal structure in the premultiplied spectrum that has been
observed experimentally \citep{Guala2006,Kunkel2006} has also been
observed here, despite suggestions that the experimental observations
were an artifact of the use of Taylor's hypothesis, which is not used
here. However, it seems likely that this two-peak structure will not
continue as the Reynolds number increases and the inner and outer
peaks in the premultiplied spectra become further separated in $k$ and
in $y$.